\documentstyle[11pt,aaspp4]{article}

\newcommand{\masr}{mas yr$^{-1}$}

\newcommand{\uas}{$\mu$as}
\newcommand{\cdeg}{^\circ}
\newcommand{\jb}{Jy beam$^{-1}$}
\newcommand{\kms}{km s$^{-1}$}
\newcommand{\sqt}{$\sqrt{2}$}

\lefthead{Walker et al.}
\righthead{VLBA Observations of the Accretion Region in NGC~1275}

\begin{document}


\title{~~\vspace{2.0 cm}\\
VLBA ABSORPTION IMAGING OF IONIZED GAS \\ASSOCIATED WITH THE
ACCRETION DISK IN NGC~1275}

\author{R. C. Walker, V. Dhawan, and J. D. Romney}
\affil{National Radio Astronomy Observatory, Socorro, NM 87801}
\authoremail{cwalker@nrao.edu, vdhawan@nrao.edu, jromney@nrao.edu, 
jbenson@nrao.edu}

\author{K. I. Kellermann}
\affil{National Radio Astronomy Observatory, Charlottesville, VA 22901}
\authoremail{kkellerm@nrao.edu}

\author{R. C. Vermeulen}
\affil{Netherlands Foundation for Radio Astronomy, Dwingeloo, The Netherlands}
\authoremail{rcv@nfra.nl}

\vspace{1 cm}
\begin{center}
{Accepted 1999 Sept. 9 for publication in the Astrophysical Journal.}
\end{center}

\begin{abstract}

Nearly simultaneous VLBA observations of 3C~84, the radio source
associated with NGC~1275, have been made at multiple frequencies to
study the free-free absorption of the northern, or ``counterjet'',
feature found by Walker, Romney, and Benson \markcite{W94}(1994) and
by Vermeulen, Readhead, and Backer \markcite{V94} (1994).  Our
observations confirm that the spectra are consistent with free-free
absorption and eliminate the possibility that the earlier result was
an effect of variability.  The northern feature is well resolved
spatially, so images have been made showing the distribution of the
absorption over a region of about 1.5 pc on a side, beginning about
1.5 pc from the presumed location of the central object.  That
distribution is dominated by a strong decrease with radial distance.
The magnitude of the absoption near 2.5 pc projected distance from the
central object is consistent with a $10^4$ K gas with an emission
measure of about $5 \times 10^8$ pc cm$^{-6}$.  The geometry is
consistent with absorption by ionized gas associated with an accretion
disk.  The data provide firm constraints for models of the outer
regions of accretion disks and, perhaps, associated winds.

\end{abstract}

\clearpage

\section{Introduction}

The dominant member of the Perseus Cluster is the galaxy NGC~1275,
which has a Seyfert-like nucleus, but an unusual morphology.
Minkowski \markcite{M57}(1957) found that it contains two velocity
systems, one at 5300 \kms and the other at 8200 \kms.  The low
velocity system includes the main body of NGC~1275 (Strauss et
al. \markcite{S92}1992) and filaments of ionized gas thought to be a
product of the cooling flow that is centered on the galaxy.  CO and
X-ray observations suggest the cooling flow rate is several hundred
${\rm M}_{\sun} {\rm yr}^{-1}$ (Lazareff et al. \markcite{L89}1989;
Allen \& Fabian \markcite{A97}1997).  HST and Keck observations of
globular clusters suggest that many may have been formed in a merger
event several hundred million years ago (Carlson et
al. \markcite{C98}1998; Brodie et al. \markcite{B98}1998).  The high
velocity system includes giant H~II clouds.  Hydrogen associated with
the high velocity system is seen in absorption against the nuclear
radio emission (DeYoung, Roberts, \& Saslaw \markcite{D73}1973; Romney
\markcite{R78}1978).  This and other evidence indicate that the high
velocity system is either infalling, or is a system that is already
part way through the main galaxy (e.g. Kaisler et
al. \markcite{K96}1996; N{\o}rgaard-Nielsen et
al. \markcite{N93}1993).

The radio source 3C~84 is associated with NGC~1275.  The features
coincident with the center of the galaxy constitute one of the
strongest compact radio sources in the sky.  This source began a major
increase in activity in about 1959.  It was below 10 Jy in the
earliest observations, but was already rising towards peaks in excess
of 50 Jy in the 1970's and 1980's (Dent \markcite{D66}1966;
Pauliny-Toth \& Kellermann \markcite{P66}1966; Pauliny-Toth et
al. \markcite{P76}1976; O'Dea, Dent, and Balonek \markcite{O84}1984;
Nesterov, Lyuty, \& Valtaoja \markcite{N95}1995).  It has been high
ever since, although in recent years the flux density had been
decaying and is currently around 20 Jy at centimeter wavelengths.  

The compact structure of 3C~84 has been the object of Very Long
Baseline Interferometry (VLBI) observations since the early days of
that technique (See, for example, Pauliny-Toth et
al. \markcite{P76}1976; Readhead et al. \markcite{R83}1983; Romney et
al. \markcite{R84}1984; Marr et al. \markcite{M89}1989; Biretta,
Bartel, \& Deng \markcite{B91}1991; Krichbaum et
al. \markcite{K93}1993; Venturi et al. \markcite{V93}1993; Dhawan,
Kellermann, and Romney \markcite{D98}1998; Romney, Kellermann, \&
Alef \markcite{R99}1999).  It has complex structure on parsec scales
that frustrated imaging efforts in the early years.  But early models,
combined with more recent images made with the global networks and
with the Very Long Baseline Array (VLBA --- Napier et
al. \markcite{N94}1994) show that, south of the compact core, there is
a bright region of emission that has been increasing in length by
about 0.3 milliarcseconds yr$^{-1}$ (\masr) and whose size
extrapolates back to zero at about the time of the 1959 outburst
(Romney et al. \markcite{R82}1982; the rate is from Walker, Romney, and
Benson \markcite{W94} 1994, hereafter WRB, and is based on their image
and others from the literature).  The bright region now extends about
15 mas south of the compact core.  It is almost certainly related to a
jet, but the morphology is somewhat like the radio lobes seen in many
sources on much larger scales.  This may indicate that the jet
material from the 1959 event is not simply following earlier material
down the jet channel, but is interacting strongly with either that
material or the surrounding medium.  To avoid a physical bias, we will
simply call the bright regions ``features''.  For this study, they are
only used as sources of background radiation with which to study the
absorbing medium.

Beyond the bright southern feature, the remnants of earlier activity
are clearly seen.  Immediately south of the 15 mas bright feature, low
frequency VLBI observations show a continuation of the jet to about
100 mas (e.g., Taylor \& Vermeulen \markcite{T96}1996; Silver,
Taylor, \& Vermeulen \markcite{S98}1998, hereafter STV; this paper).
STV also see a ``millihalo'' approximately 250 mas in size at 330 MHz.
On scales of arcseconds and larger, 3C~84 has weak, somewhat asymmetric
twin jet structure surrounded by a diffuse halo (Pedlar et
al. \markcite{P90}1990; Sijbring \markcite{S93}1993).

In 1993, a feature extending to about 8 mas north of the compact core
was discovered with the VLBA at 8.4 GHz by WRB and independently,
using 1991 Global VLBI Network data at 22 GHz, by Vermeulen, Readhead,
and Backer \markcite{V94}(1994, hereafter VRB).  At 22 GHz, this
feature was similar to, but smaller and somewhat weaker than the
bright southern feature.  It most likely is related to a jet on the
far side of the AGN corresponding to the near side jet responsible for
the southern feature.  This source provides an especially favorable
case for relating jet and counterjet features because of the dramatic
changes in jet brightness dating to the 1959 outburst.  WRB show that
the ratio of distances to the ends of the bright emission in the north
and south, the brightness ratio of the features at 22 GHz, and the
measured speed of the southern feature, are consistent with a mildly
relativistic jet ($0.3c$ to $0.5c$) at a moderate angle ($30\cdeg$ to
$55\cdeg$) to the line-of-sight.

The most exciting result of WRB and VRB was the observation that the
northern feature has a strongly inverted spectral index.
It was very much
brighter at 22 GHz than at 8.4 GHz, and it had not been seen at all in
high dynamic range images of Biretta et al. \markcite{B91}(1991) at
1.7 GHz.  But the 22 and 8.4 GHz observations were separated in time
by about 2 years so there was some potential for confusion due to time
dependent effects.  VRB present persuasive arguments that the spectral
index, if confirmed, is most likely the result of free-free absorption
by ionized material along the line-of-sight to the northern feature.
This absorbing material must be located in the inner few parsecs of
the source --- otherwise it is highly unlikely that it would absorb
only the northern feature leaving the core and southern feature
unaffected.  An obvious geometry consistent with the data puts the
ionized gas in, or associated with, a disk.  Levinson, Laor, and
Vermeulen \markcite{V95}(1995, hereafter LLV) discuss models involving
accretion disks at some length.  One problem is that, according to
simple theory, the accretion disk should be too cold at parsec
distances to contain ionized gas.  They address a variety of
mechanisms that would allow ionizing radiation from the central
regions of the AGN to reach the outer disk and create an ionized
atmosphere adequate to explain the observed absorption.

Apparent free-free absorption is also seen in this source at lower
frequencies.  O'Dea, Dent, and Balonek \markcite{O84}(1984) deduce,
from flux density monitoring data, that the variable core component is
absorbed below about 2.2 GHz.  They suggest that the absorbing gas has
a thickness of $1.5 < L < 5$ pc, $T \sim 10^4$ K, and $n \sim 2 \times
10^3$ cm$^{-3}$.  More recently, STV found a feature 80 mas north of
the core in 330, 612, and 1414 MHz VLBA observations that presumably
corresponds to an earlier epoch of activity.  The spectrum of this
feature can be interpreted in terms of free-free absorption on a much
larger scale.

Evidence for free-free absorption on parsec scales has been seen in a
few other sources.  Jones et al. \markcite{J96}(1996) find
preferential absorption of counterjet emission in Cen~A, indicating a
geometry similar to what is seen in 3C~84.  Jones and Wehrle
\markcite{J97}(1997) see a feature in NGC~4261 that is possibly due to
absorption by an inner accretion disk with a width of less than 0.1
pc, but they do not have spectral information on this feature.
Kellermann, Vermeulen, Cohen, and Zensus \markcite{K99}(1999) report
the presence of free-free absorption of the nucleus and receding jet
in NGC~1052, with the absorption moving to lower frequencies over
scales from 15 light days to more than 3 light years.  Taylor
\markcite{T96}(1996) attributes the spectral turnover seen in the
parsec scale jet components in Hydra~A to free-free absorption over
the entire central region, probably due to gas on a somewhat larger
scale.  Parsec scale ionized gas associated with a disk may have been
observed directly in NGC1068 by Gallimore, Baum, \& O'Dea
\markcite{G97}(1997).  The inferred temperature of the observed region
is around $10^{6.7}$K.  But the case that this is ionized disk gas,
rather than a jet, is not firmly established.

Ulvestad, Wrobel, \& Carilli \markcite{U99}(1999) report evidence for
absorption in Markarian 231, a powerful Seyfert~1/starburst galaxy at
$z=0.0422$ whose compact emission regions look remarkably similar to
3C~84.  In both magnitude and physical scale, the observed absorption
in Mrk~231 is more comparable to that observed in 3C~84 by STV than
that reported in this paper.  Unlike 3C~84, a spectral gradient is
reported for the southern feature, implying some absorption over the
whole source.  But this may be a result of alignment of the images on
the core.  At least in 3C~84, the centroid of the core shifts with
frequency.  Also, the 1.4 GHz image shows emission well north of any
seen at higher frequencies, implying unphysically steep spectral
indices or some imaging effect such as much greater sensitivity
to large structures at 1.4 GHz than at the other frequencies.

In this paper, we present results of the first two epochs of
observations of 3C~84 designed to study the absorption and its two
dimensional distribution.  The observations were made with the Very
Long Baseline Array of the National Radio Astronomy Observatory,
supplemented in some cases by a single antenna of the Very Large Array
(VLA).  At each epoch, a wide range of frequencies were observed, all
within a period short enough to preclude uncertainties due to
variability.  High dynamic range images, all convolved to the same
resolution, allow us to image the absorption over the extent of the
northern feature.  

The first, and simplest, result is that the apparent absorption is not
due to temporal changes in the source.  The quasi-instantaneous radio
spectra at positions on the northern feature are consistent with
free-free absorption.  Minor deviations from a theoretical free-free
spectrum are probably due to inhomogeneities in the medium.  We show
that the dominant spatial structure of the absorption is a strong
decrease with radial distance from the central feature.  The concept
that the ionized gas is associated with the accretion disk and falls
off in density from the center is consistent with the observations.

At $v=5300$ \kms, $1\ {\rm mas} = 0.25\ h^{-1}$ pc, where $H_o = 100\
h $ \kms Mpc$^{-1}$.  In this paper, we assume that $h = 0.75$, so the
scale of the images is very close to $1\ {\rm pc} = 3\ {\rm mas}$.

\section{The Observations}

Observations of 3C84 were made with the VLBA in 1995 January and
October at 2.3, 5.0, 8.4, 15.4, 22, and 43 GHz.  The specific dates
are listed in Table~\ref{obsdetails}.  The observations were spread
over several days at each epoch in order to gather sufficient data at
each frequency to make high dynamic range images.  However all
observations for each epoch were confined to a period of two weeks in
order to avoid confusion by possible time variations.  A single
antenna of the VLA was included in the 22 and 43 GHz observations in
October in order to enhance the coverage of short baselines.  The day
following the last of our October observations, 3C~84 was observed at
0.33, 0.61, and 1.4 GHz by STV.  Therefore near simultaneous
observations of this source in 1995 October are available at all bands
for which the VLBA has receivers.

\placetable{obsdetails}

\begin{deluxetable}{lccccc}
\tablenum{1}
\tablewidth{0pt}
\tablecaption{VLBA Observations of 3C~84 \label{obsdetails}}
\tablehead{
  \colhead{Date} & 
  \colhead{Frequency} &
  \colhead{Antennas} &
  \colhead{Bit Rate} &
  \colhead{Polarization} &
  \colhead{Baseline}  \nl
  \colhead{(1995)}& 
  \colhead{(GHz)} &
  &
  \colhead{(Mbps)} &
  &
  \colhead{Hours}
  }  

\startdata
Jan 15  &   22.2         &  VLBA       &  64 & dual & 384 \nl 
Jan 22  &    2.3 \tablenotemark{a} \tablenotemark{b}        &  VLBA \tablenotemark{c} &  32 & dual & 159 \nl 
             &    5.0         &  VLBA \tablenotemark{c} &  64 & dual &  88 \nl 
             &    8.4 \tablenotemark{b}        &  VLBA \tablenotemark{c} &  32 & dual & 158 \nl 
Jan 29  &   15.3         &  VLBA       &  64 & LCP  & 249 \nl 
Jan 30  &   43.2         &  VLBA       & 128 & LCP  & 399 \nl 
\nl					 
Oct 09  &   15.3         &  VLBA       & 128 & dual & 174 \nl 
Oct 16 \tablenotemark{d}
             &   43.2         & VLBA+VLA \tablenotemark{e}   & 128 & LCP  &  \nl 
Oct 21  &   22.2         & VLBA+VLA \tablenotemark{e}   & 128 & dual & 244 \nl 
Oct 22  &    2.3 \tablenotemark{a}        &  VLBA       &  64 & dual &  44 \nl 
             &    5.0         &  VLBA       &  64 & dual & 116 \nl 
             &    8.4         &  VLBA       &  64 & dual & 135 \nl 
Oct 23 \tablenotemark{d} \tablenotemark{f} &  
                0.3, 0.6, 1.4 &  VLBA       &  64 & dual & \nl 
\enddata
\tablenotetext{a}{Band switching was used to observe 2.3, 5.0, and 8.4 GHz on the same day.}
\tablenotetext{b}{Dichroic system used to observe 2.3 and 8.4 GHz simultaneously on Jan. 22.}
\tablenotetext{c}{Only 9 of the 10 VLBA antennas participated.}
\tablenotetext{d}{Data not used for this paper, but listed for completeness.}
\tablenotetext{e}{Included one antenna only of the VLA.}
\tablenotetext{f}{Independent but related project (Silver, Taylor, \& Vermeulen 1998).}
\end{deluxetable}

The northern feature seen at higher frequencies about 8 mas from the
core was not detectable at 2.3 GHz to a limit of about 2 m\jb\ in both
January and October, presumably because it is too heavily absorbed.
Also the much more distant northern feature, at about 80 mas, that was
reported by STV at low frequencies, was not seen in our 2.3 GHz
images, perhaps because of inadequate short baselines.  The convolved
images used to look for the distant features have off source noise
levels of about 1.5 m\jb\ with beams of 19 by 15 mas.  At 43 GHz the
the feature 8 mas north of the core was detected, but most of the flux
density appears to be missing.  This feature is several mas in size,
which is a size scale that falls in the range that is poorly measured
at 43 GHz because of the gap in (u,v) coverage between the VLBA and
the VLA.  The feature is mostly resolved out on the VLBA, but is still
a point source to the VLA.  Therefore the results presented here will
be based almost entirely on the 5.0, 8.4, 15.4, and 22 GHz images,
although we show the 2.3 GHz images and one 43 GHz image.

Details of the observations are presented in Table~\ref{obsdetails}.
The columns are the date within 1995, the antennas included, the bit
rate, the polarizations recorded, and the number of effective baseline
hours of data available for the final image.  The total recorded bit
rate was divided among 2, 4, or 8 baseband channels.  The baseline
hours listed is simply the integration time times the number of
visibility records.  The 2.3, 5.0, and 8.4 GHz observations were
performed by band switching during a single day in both January and
October.  In January, but not in October, the 2.3 and 8.4 GHz data
were obtained simultaneously using the VLBA's dichroic system.

Apriori amplitude calibration was done in the usual way using the
system temperatures measured during the observations and gains
measured by the NRAO staff during pointing observations.  Most of the
data were processed with AIPS.  The 22 GHz image from January was made
using DIFMAP (Shepherd, Pearson, \& Taylor \markcite{S94}1994).  On
1995 October 20, VLA observations were made of 3C~84 and the compact
source DA~193 (0552+398) to obtain flux densities.  These were used to
check and adjust the apriori amplitude calibration of the VLBA.
Adjustments in the amplitude scales by factors of 1.00, 0.91, 1.00,
0.95 for 22, 15.4, 8.4, and 5.0 GHz were made as a result.  The
remaining uncertainty in the flux scale is roughly estimated to be
about 5\% for the October data.  For January, the apriori calibration
had to be trusted and there may be residual errors at the 10\% level.
The 2.3 GHz amplitudes from January were scaled by 0.87 to adjust for
the effects of an inappropriate off-source position used in VLBA
2.3/8.4 GHz dual frequency gain measurements prior to 1997 November.

\begin{figure}[t]
\plotfiddle{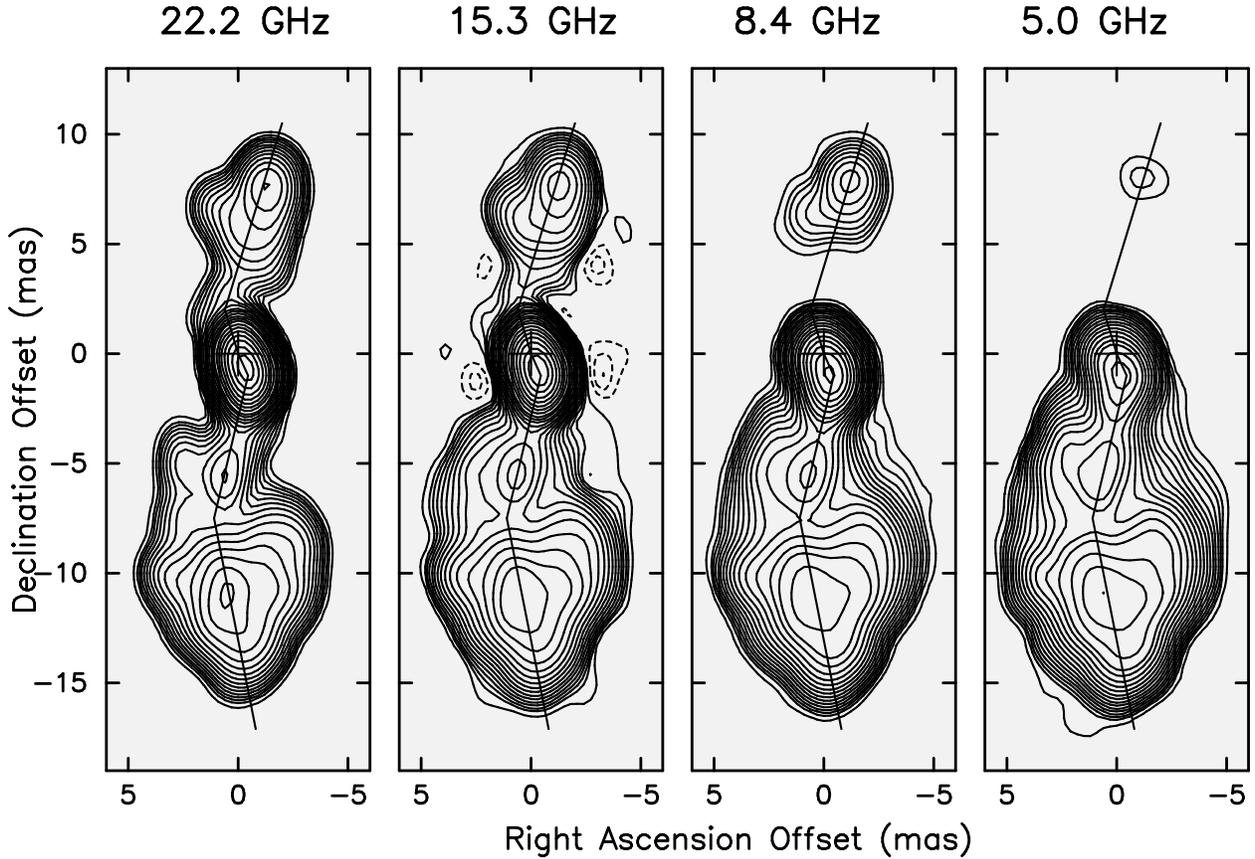}{4.5in}{-90}{73}{73}{-294}{390}
\caption{Montage of the VLBA images of 3C~84 based on data
taken in 1995 January.  These displays are based on the CLEAN
components convolved with a common Gaussian beam of 1.6 by 1.2 mas,
elongated north-south.  The contour levels start with 5, 10, 14, 20
m\jb\ and increase from there by factors of \sqt.  The residual image
is not included, but the CLEANs are sufficiently deep, and the lowest
contour sufficiently high, that this does not have a significant
effect.  The north-south segmented line shows the location of the
slice along which some of the analysis was done.  Note that, for $H_o
= 75$ \kms Mpc$^{-1}$, the scale is about 3 mas per
parsec. \label{montageJ}}
\end{figure}

\placefigure{montageJ}

\begin{figure}[t]
\epsscale{0.90}
\plotfiddle{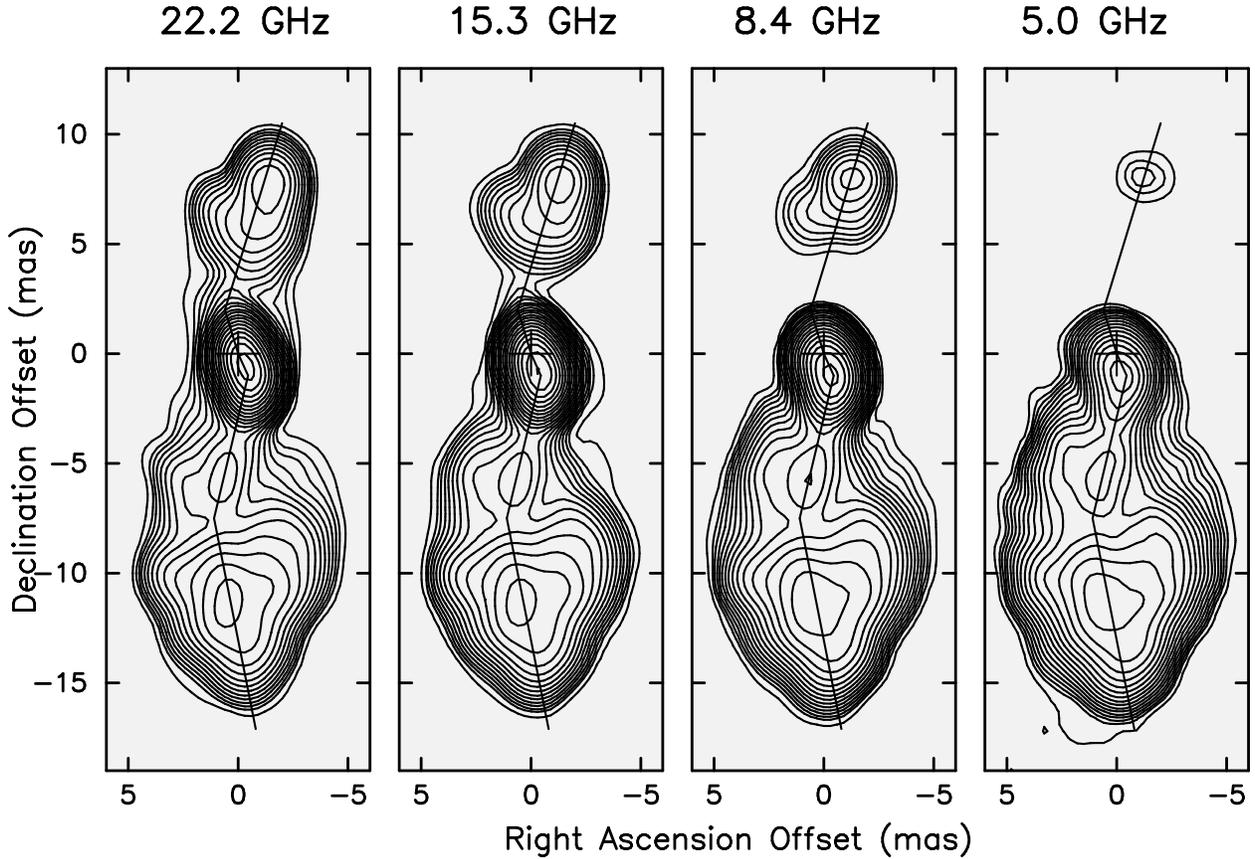}{4.5in}{-90}{73}{73}{-294}{390}
\caption{Montage of the VLBA images of 3C~84 based on
data taken in 1995 October.  The beam, contour levels and slice are the
same as for Figure~\ref{montageJ}. \label{montageO}}
\end{figure}

\placefigure{montageO}

Montages of the contour plots of the images used in the analysis are
shown in Figures~\ref{montageJ} and \ref{montageO}.  All images are
convolved to a beam of 1.6 by 1.2 mas, elongated in position angle 0
degrees.  This is about the actual resolution of the 5 GHz images, but
represents a considerable smoothing of the higher frequency images.
On each contour plot, a segmented line is shown which marks the slice
used for some of the analysis below.  

\begin{figure}[thp]
\epsscale{1.0}
\plotone{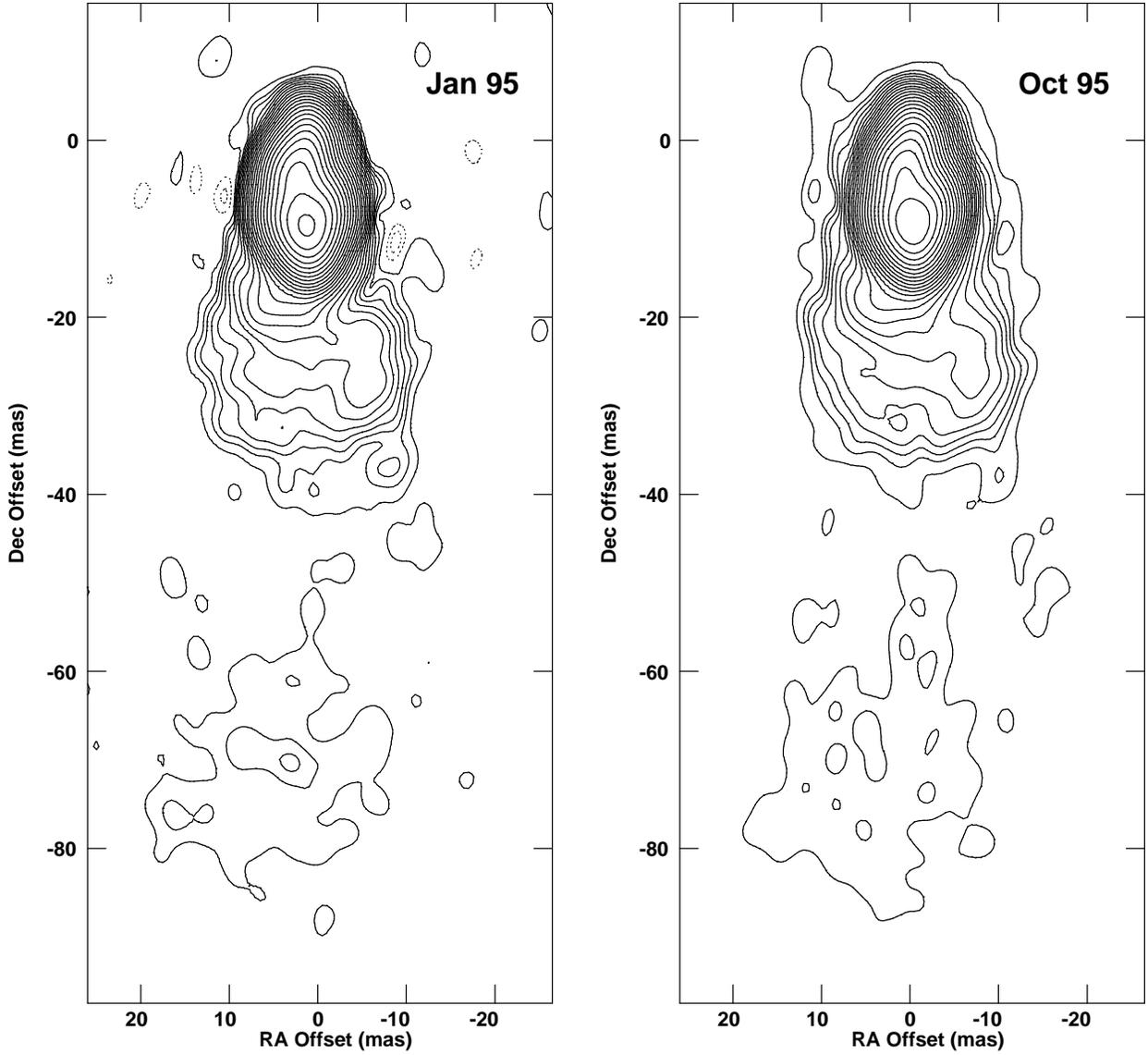}
\figcaption[f3.eps]{The images of 3C~84 at 2.3 GHz from 1995 January
and 1995 October.  The convolving beams are $4.4 \times 3.2$ mas
elongated north-south.  These are fully restored CLEAN images --- they
include the residuals, unlike the images in Figures~\ref{montageJ} and
\ref{montageO}.  The contours are 2, 4, 5.7, 8, m\jb\ increasing from
there by factors of the \sqt.  The peaks in the images are
8.84 \jb\ in January and 7.90 \jb\ in October.  No effort has been made
to carefully align these images either with each other or with those
of Figures~\ref{montageJ} and \ref{montageO}.  Note the scales ---
these images show emission well south of what is seen at higher
frequencies.  There clearly have been previous outbursts of this
source.  
\label{sband}}
\end{figure}

\placefigure{sband}

Contour plots of the 2.3 GHz images are shown in Figure~\ref{sband}.
The resolution of these images is lower than for the others because of
the lower frequency.  These images show the southern feature well and
make it clear that there is a weak, underlying structure extending
well beyond the region of the features that resulted from the 1959
outburst.  Those underlying structures are also seen weakly in the 5
GHz images, although only when plotted to deeper levels than were used
in Figures~\ref{montageJ} and \ref{montageO}.  The northern feature is
absent, presumably completely absorbed.  In the 1995 October image at
22 GHz, the ratio of integrated flux density in the northern and
southern features is about 3.7.  In the 2.3 GHz image from the same
epoch, the integrated flux density of the southern feature is about 21
Jy.  If the intrinsic spectral index (without absorption) of the
northern and southern features is the same, the expected flux density
of the northern feature at 2.3 GHz would be 5.7 Jy.  Instead it is
below about 2 mJy, so there is very strong absorption ---
corresponding to an optical depth of at least 7.9.  At 5 GHz, the
northern feature had an integrated flux density of about 20 mJy in
January and 34 mJy in October in a region smaller than the 2.3 GHz
beam.  Thus the observed spectral index between 2.3 and 5.0 GHz was at
least 2.9 in January and 3.6 in October.  While we do not use this
information in the following analysis, it further supports the
interpretation of the spectra as being due to the exponential cutoff
resulting from free-free absorption toward the northern feature.

\begin{figure}[t]
\epsscale{0.4}
\plotone{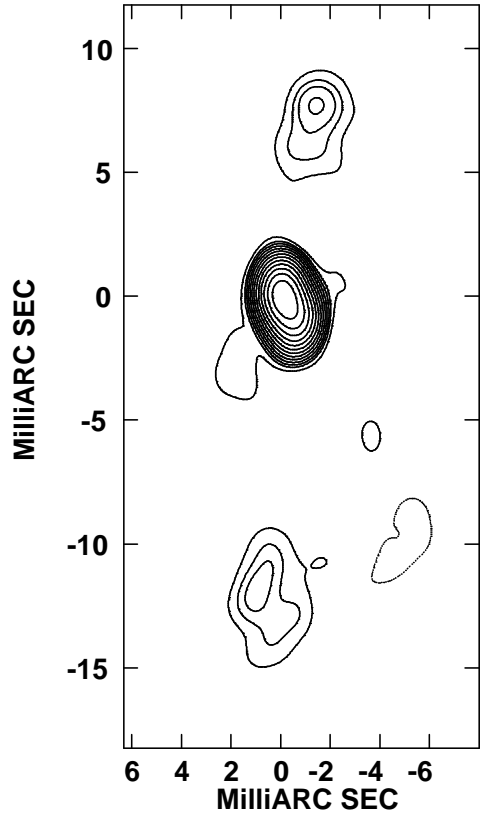}
\figcaption[f4.eps]{This image shows the results at 43 GHz from
1995 January, convolved to the same resolution as
Figures~\ref{montageJ} and \ref{montageO}.  The lowest contour levels
are 20 and 40 m\jb, increasing by factors of \sqt\ from there.  It
is clear that most of the flux density from the northern and southern
features is missing in these images.  These structures fall in a size
range that, at 43 GHz, is still unresolved by the VLA but is almost
completely resolved out by the shortest VLBA baselines.  Therefore,
this image is only used to demonstrate that the features are present
at 43 GHz, but not to place any constraints on them.
\label{qband}}
\end{figure}

\placefigure{qband}
\notetoeditor{This figure should be small}

An image based on the 43 GHz VLBA data from Jan. 1995 is shown in
Figure~\ref{qband}.  The image has been convolved to the same beam
used in the analysis of the 5 to 22 GHz images.  The northern and
southern features are clearly there, but at much reduced flux
densities because of the lack of baselines of a few tens to a few
hundred km.  Because of this missing flux density, this image cannot
be used in the analysis.  For more on VLBA monitoring of the 43 GHz
structure of this source, see Dhawan, Kellermann, and Romney 
\markcite{D98}(1998).

\begin{figure}[t]
\plotfiddle{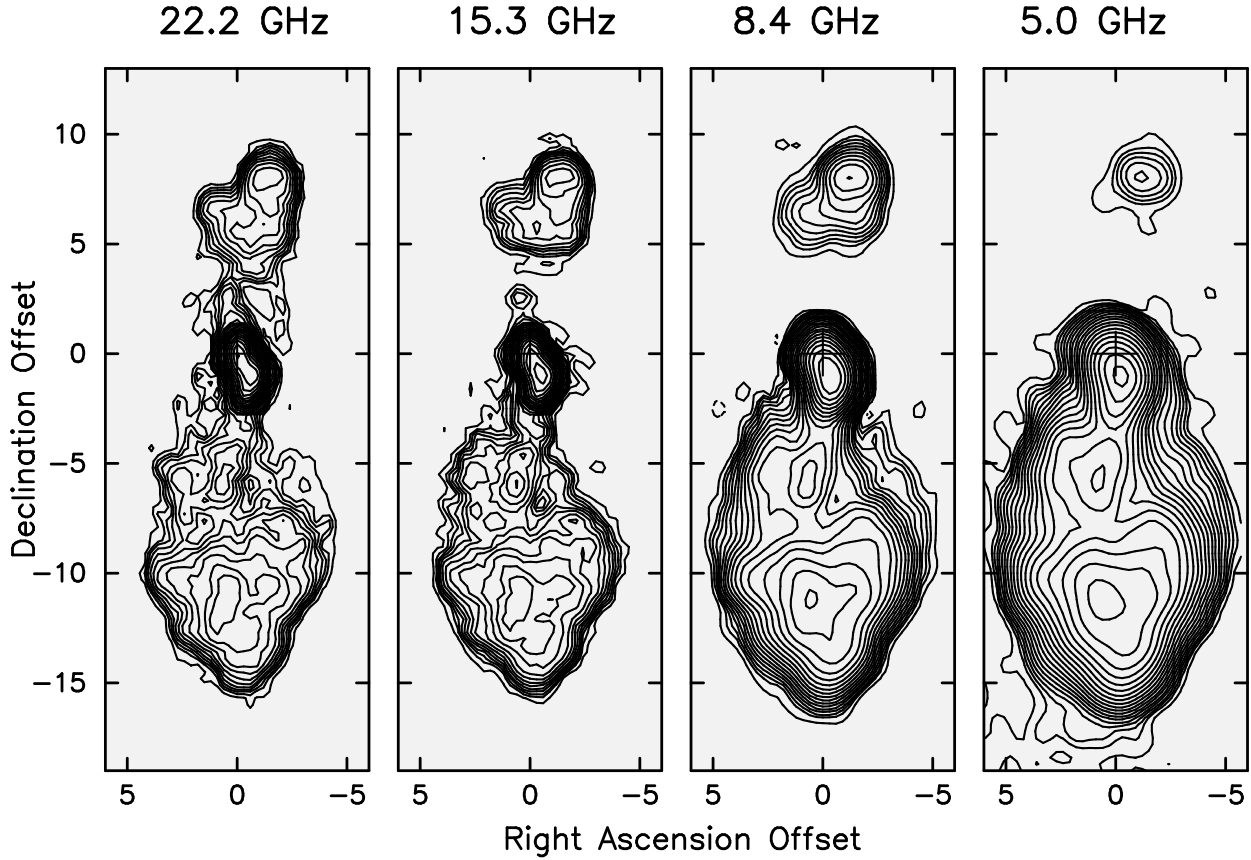}{4.5in}{-90}{73}{73}{-294}{390}
\figcaption[f5.eps]{Montage of full resolution images of 3C~84 from
1995 October.  The displays are based on the same CLEAN components as
were used for Figure~\ref{montageO}, but using the full resolution of
the original images.  The convolving beams are (major axis in mas,
minor axis in mas, position angle in degrees) $(0.73, 0.48, -2.90)$
for 22 GHz, $(0.68, 0.50, 2.93)$ for 15 GHz, $(1.21, 0.88, -2.61)$ for
8.4 GHz, and $(1.6, 1.2, 0.0)$ for 5.0 GHz.  The lowest contour levels
are 2 and 4 m\jb, increasing by factors of \sqt\ from there.  The
beams, especially the 22 GHz beam, don't scale with frequency because
different tapers and weighting schemes were used.  \label{fullO}}
\end{figure}

\placefigure{fullO}

For background information, full resolution images of the source in
1995 October are shown in Figure~\ref{fullO}.  The analysis in this
paper will be based entirely on the images convolved to the resolution
of the 5 GHz images.  At higher resolutions, imaging artifacts, that
are not readily visible in the total intensity images, become apparent
in spectral index images and in the absorption analysis, which are
based essentially on differences.  These artifacts are the result of
the inadequate (u,v) coverage of the VLBA at high frequencies for a
source of this complexity and size.  Future observations will include
additional antennas to address this difficulty.  The 22 GHz image was
not made with the instrument's full resolution, knowing that it was to
be convolved to the lower resolution of the lower frequency
observations.  Note that the 5 GHz image, with contour levels that are
deeper than those of Figure~\ref{montageO}, shows the beginning of the
extension, beyond 15 mas south of the core, that is seen at 2.3 GHz.

\section{Analysis}

The jet speed and angle results of WRB are based on the measured
length ratio of the northern and southern features, on the long term
expansion rate of the southern feature, and on the assumption that the
northern and southern features both originated in the outburst of
about 1959, so they have the same age.  The 1995 data do not change
the numbers.  But the flux density ratio of the northern and
southern features at 22 GHz was 9 in the data from VRB, used by WRB.
In the 1995 October 22 GHz image, the ratio is about 3.7.  The
difference might be the result of the better coverage of short
interferometer baselines in the 1995 data, or it might be the result
of real changes in either the emission in the source or in the
absorbing screen suggested by VRB and discussed below.  By the beamed
jet model of Blandford and K\"onigl \markcite{B79}(1979), the flux
density ratio $R$ and length ratio $D$ are related by $R=D^{\eta}$.
The index $\eta$ is either $(2-\alpha)$ or $(3-\alpha)$, where
$\alpha$ is the spectral index ($S\propto \nu^{\alpha}$), 
for a continuous jet or a single
component respectively.  The data of VRB give $\eta=3.7$, which is
consistent with the single component model and the spectral index
observed in the southern feature.  The 1995 data give $\eta = 2.2$,
which does not fit the simple model so well.  Even with the continuous
jet model, the spectral index would have to be flatter than is
observed in the southern feature.  But the flux density of the source
has been decreasing with time for many years.  In any one image, the
northern feature is observed at an earlier time in its history than
the southern feature because of light travel time effects --- about 22
years earlier for the geometry of WRB and our assumed Hubble constant.
In the early to mid 1970's, the integrated flux density of 3C~84 at 22
GHz was roughly twice the the 1995 value (Nesterov, Lyuty, \& Valtaoja
\markcite{N95}1995).  Therefore, while the distribution of the
emission at 22 GHz in the 1970's is not known, it is reasonable that
the northern feature would be stronger ($\eta$ lower) than would be
expected by the simple, no-evolution beaming model.

The southern feature has a fairly uniform spectral index of $\alpha
\approx -0.7$, a value rather typical of radio jets and hot spots in
lobes.  The northern feature is entirely different.  It is far
stronger at high frequencies and has spectral indices that vary both
spatially and with frequency and that get as high as $\alpha > +4$.
Even with less data, VRB argued that the steeply inverted spectral
index, along with the extended size of the feature, effectively rules
out other explanations than free-free absorption.  Below we
characterize and discuss this absorption.  The core region also has an
inverted spectrum and may be subject to some free-free absorption.
However that spectrum could also simply be due to synchrotron
self-absorption in the most compact regions of the jets.  Therefore we
do not include the core region in our discussion of free-free absorption.

Most of the analysis for this study was done with a special purpose
program that operates on the CLEAN components from the AIPS images.
Very deep CLEANs were done, so essentially all of the flux density is
represented in the CLEAN components.  The program is used to shift the
components to align the images, make any required final amplitude
calibration adjustments, convolve the components with a Gaussian beam
roughly equivalent to the resolution of the lowest frequency image,
and fit for the absorption.  The fits are done both in the image plane
and for points along a segmented slice as shown in
Figures~\ref{montageJ} and \ref{montageO}.

In order to image the absorption, it is necessary to align the images
at different frequencies very accurately.  Because of the presence of
steep emission gradients, relative shifts of as small as 20
microarcseconds (\uas) make small, but perceptible, changes in the
distribution of the absorption.  An alignment to that level of
accuracy is desirable, but VLBI observations that don't utilize phase
referencing retain no absolute position information.  These
observations were scheduled with occasional short scans on the nearby
calibrator 0309+411.  Positions accurate to about a 1 mas could probably
be obtained with the aid of these scans.  But to obtain positions to
20 \uas\ would require dual frequency observations to remove the
effects of the ionosphere and scans all over the sky to calibrate the
troposphere.

It is possible to use the source itself, in the case of 3C~84, to
align the images very accurately.  This requires the assumption that
the southern feature is similar in structure at all frequencies ---
that there aren't large variations in observed spectral index.  This
assumption is supported by the images.  The images could be aligned by
finding the relative positions that minimize spectral index gradients.
But a method that is at least as effective, and easier to automate, is
to utilize the image cross correlation techniques previously developed
for use on VLA data on 3C120 and M87 (Walker \markcite{W97b}1997 and
references therein).  The idea is to find the peak, as a function of
position offset, of the cross correlation of portions of two images.
For comparison, the software also does a least squares fit for
the position offset of the image portions, generally obtaining similar
results.  For 3C~84, the regions greater than 5 mas south of the core
were cross-correlated for all 6 possible pairs of images at each
epoch.  The images were shifted so as to minimize the offsets
determined by the cross correlations.  The 6 pairs over determine the
problem, so the scatter of the results gives a crude estimate of the
accuracy of the method.  For each epoch, an alignment was found for
which all reported offsets in RA and Dec were less than 5 \uas, easily
meeting our requirements.  With these alignments, the uniformity of
spectral indices was better than had been achieved previously with a
tedious, non-automated procedure based directly on the spectral
indices.

With alignment nominally accurate to a few microarcseconds, the
internal motions of the source start to be a concern.  Historically,
the southern lobe is moving way from the core at about 0.3 \masr.  The
counterjet would be expected to be moving at some fraction of that
speed, probably lower by the length ratio of about 1.8.  Thus we
expect relative motions of jet and counterjet at the level of about
1.3 \uas\ per day.  The January observing sequence spanned 15 days and
the October sequence spanned 13 days, so relative motions of up to 19
\uas\ are expected.  This has very little effect on the final results.
Nevertheless an attempt has been made to account for it by moving
the features more than 2 mas north of the core by an appropriate
amount, depending on observing date relative to the day of the 5 and
8.4 GHz observing.  Rather than having an abrupt transition between
the shifted and unshifted components, the region between -3.6 mas and
+2 mas north of the core was stretched linearly.  These distances are
in proportion to the jet/counterjet length ratio.

The analysis program is capable of doing simultaneous fits for 22 GHz
amplitude ($I_o$), emitted spectral index ($\alpha_e$), absorption, and a
covering factor ($f$), all as a function of position.  The equation of 
the fit is:
\begin{equation}
 I_\nu = I_o \times \left(\nu \over{ 2.2 \times 10^{10}} \right)^{\alpha_e}
   \left[ (1-f) + f e^{-\kappa} \right]
\end{equation}
where $I_\nu$ is the flux density measured at frequency $\nu$ (Hz),
and $\kappa$ is the absorption coefficient, which is 
the main item of interest.  The absorption is given by: 
\begin{equation}
\kappa = 9.8 \times 10^{-3} L_{pc} ~n_e^2~ T^{-1.5}~ \nu^{-2} 
\left[17.7 + \ln( T^{1.5}~ \nu^{-1}) \right]
\label{kap}
\end{equation}
where $L_{pc}$ is the path length through the absorbing medium in
parsecs, $n_e$ is the electron density in cm$^{-3}$, $T$ is the temperature in
degrees K, and the terms in the square brackets are the Gaunt factor.
The fits for absorption were actually done for the combination of
parameters ($L_{pc} ~n_e^2~ T_4^{-1.5}g_4$), where $T_4$ is the
temperature in units of $10^4$ K and $g_4$ is the ratio of the Gaunt
factor to the Gaunt factor for $T=10^4$ K.  Note that $L_{pc}n_e^2$ is
simply the emission measure (EM), so for the case of a $10^4$ K
constant temperature medium, the fit is simply for the emission
measure.
 
\begin{figure}[tp]
\epsscale{0.7}
\plotone{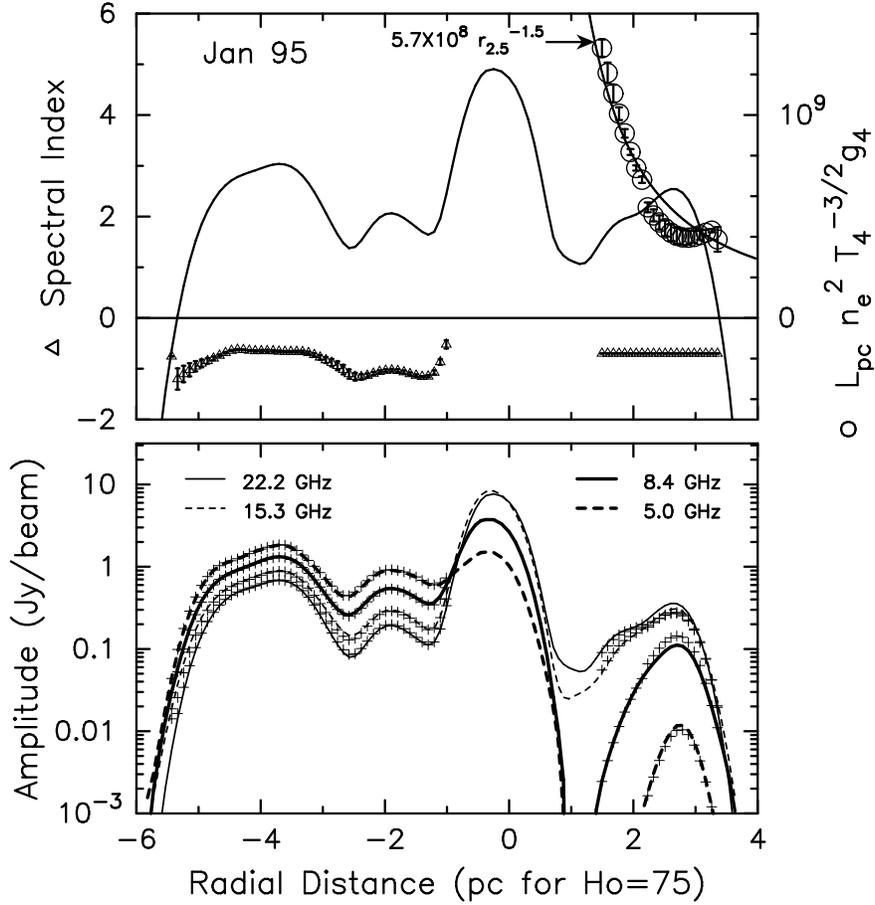}
\figcaption[f6.eps]{The results of least squares fits for emitted
spectral index and absorption along the slice shown in
Figure~\ref{montageJ}.  The horizontal axis is the radial distance
from the core (ie, not exactly along the slice) in parsecs assuming
that $h = 0.75$, for which 1 pc is about 3 mas.  The line in the top
panel is the profile of flux density per beam at 22 GHz while the
lines in the bottom panel are the flux densities at 5.0, 8.4, 15.4,
and 22 GHz, as labeled.  Note the logarithmic flux density scale.  The
triangles in the top plot show the emitted spectral index, which is
the result of a least squares fit for negative core distances (south)
and is held fixed at $\alpha_e = -0.7$ to the north.  The circles are
the fitted absorption, parameterized as described in the text.  Note
that this parameterization becomes the emission measure if $T=
10^4$~K.  The absorption is only fit to the north of the core --- to
the south there is no evidence for deviations from a simple spectral
index.  The formal error bars of the fits are shown.  The curved line
through the absorption results, and the equation at the top center of
the plot, are the result of a fit of a power law to the absorption
data.  In the lower panel, the crosses are the data as predicted from
the fit result at each point in the slice.  Note that the step at
about 1.8 pc is at the edge of the region in which it was possible to
use the 5.0 GHz data in the fits.  \label{sliceJ}}
\end{figure}

\begin{figure}[tbh]
\epsscale{0.7}
\plotone{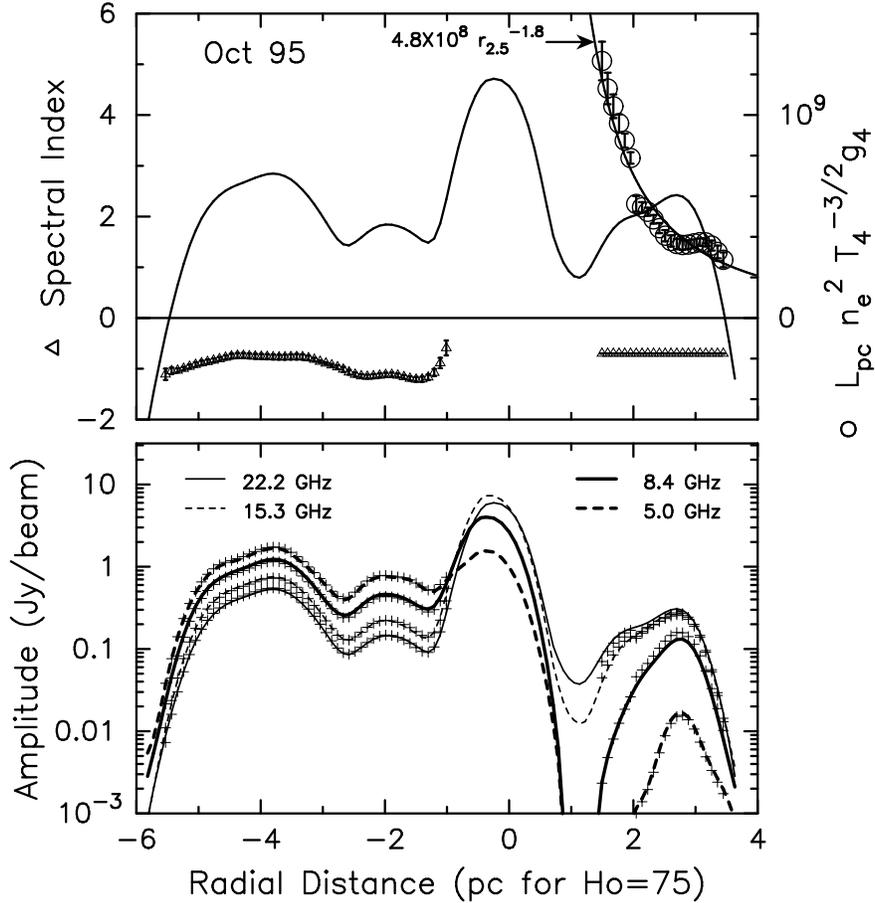}
\figcaption[f7.eps]{This is the same type of plot as
Figure~\ref{sliceJ}, but for the slice shown in Figure~\ref{montageO}.
Note the slightly lower absorption and corresponding somewhat higher
flux density on the counter feature at low frequencies.
\label{sliceO}}
\end{figure}

\begin{figure}[t]
\epsscale{0.55}
\plotone{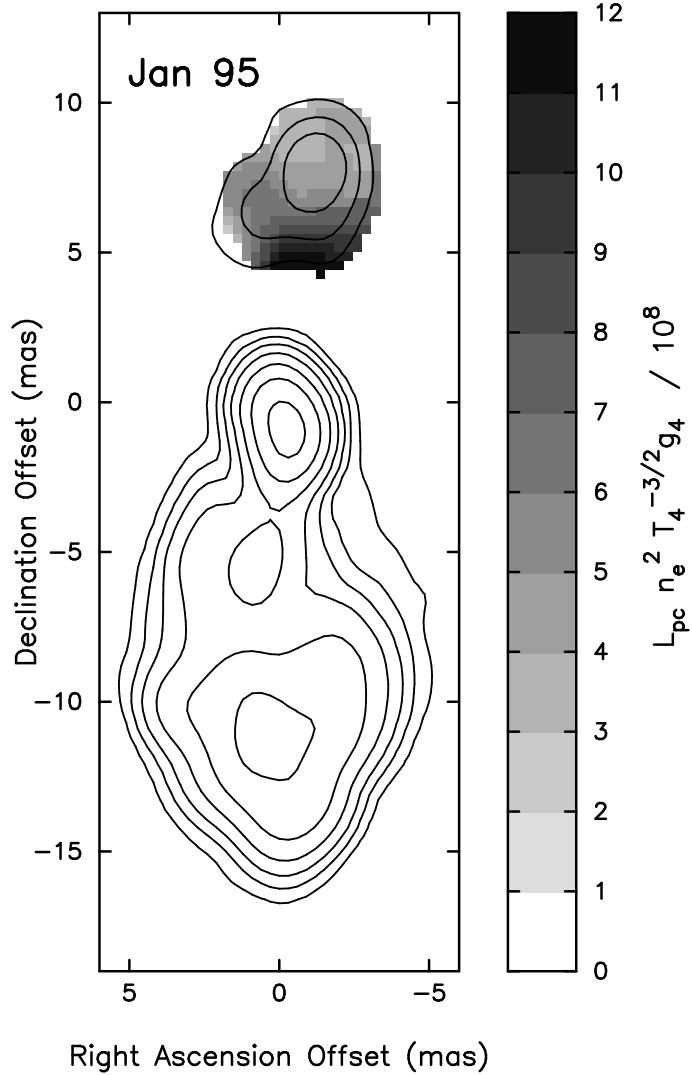}
\figcaption[f8.eps]{This image shows the two dimensional
distribution of the free-free absorption in 3C~84 over the region of
the counter feature in the 1995 January images.  There is no detected
absorption over the brighter southern feature.  The absorption is
parameterized in the same way as in Figure~\ref{sliceJ}.  The
overlayed contours are every third one from the 8.4 GHz image of
Figure~\ref{montageJ}.  The strong radial gradient away from the core
is apparent. \label{mapEMJ}}
\end{figure}

\begin{figure}[t]
\epsscale{0.55}
\plotone{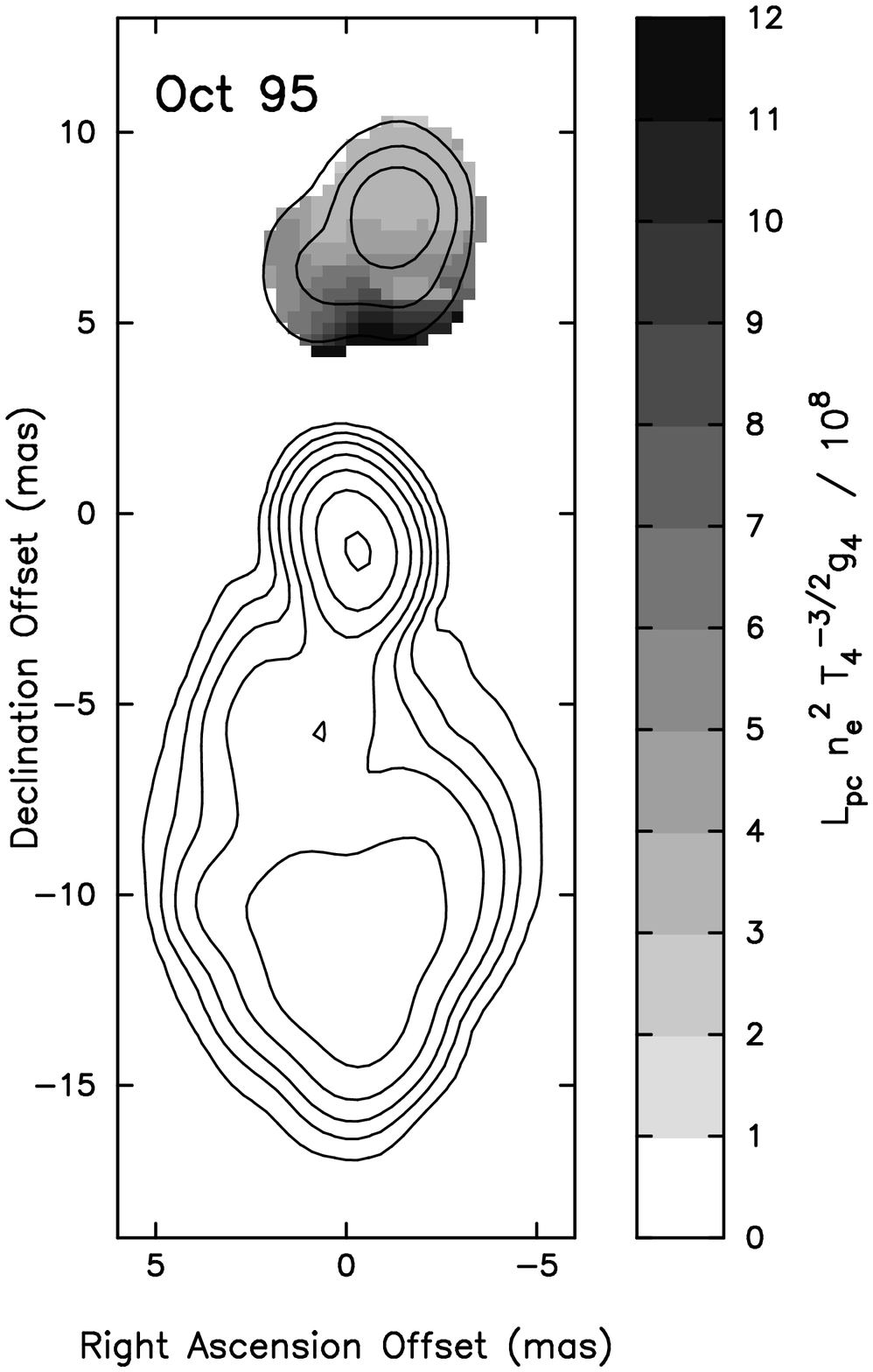}
\figcaption[f9.eps]{This is the same information as Figure~\ref{mapEMJ},
except that it is for the 1995 October data.  The time interval between
the two data sets is really too short to trust measured changes, but 
there are indications that the absorption has decreased between January
and October. \label{mapEMO} }
\end{figure}

The above parameterization has 4 unknowns.  The observations reported
here have 4 frequencies, and so have a maximum of 4 measurements at
any point in the image or along the slice --- sometimes fewer.
Fitting for all 4 parameters was not really justified.  After some
experimentation, it was decided to fit only for $I_o$ and $\alpha_e$
along the southern feature, setting $EM=0$.  On the northern
feature, the subtle variations in emitted spectral index are
overwhelmed by the strong absorption, so the fits were done for $I_o$
and absorption with $\alpha_e=-0.7$ and $f=1$.  The results for the
fits along the slice are shown in Figures~\ref{sliceJ} and
\ref{sliceO}.  The results for the absorption as a function of
position in the image are shown as the gray scales in
Figures~\ref{mapEMJ} and \ref{mapEMO}.

\placefigure{sliceJ}

\placefigure{sliceO}

\placefigure{mapEMJ}

\placefigure{mapEMO}

The quality of the fits is best assessed in Figures~\ref{sliceJ} and
\ref{sliceO}.  In those figures, the top panel shows the results for
$\alpha_e$ and absorption along the slice, in terms of radial distance
from the core.  The amplitudes from the 22 GHz slice are also
shown for orientation.  The lower panel shows the amplitudes from all
4 frequencies as lines and the fitted model amplitudes as crosses.
Note the logarithmic scale and the very deep absorption, especially at
5 GHz.  The fits are not perfect, in the sense that the absorption is
not actually quite as sharp as for free-free absorption in a uniform
medium.  Better, even near perfect, fits can be obtained by releasing
either $\alpha_e$ or $f$.  Fitting for $\alpha_e$ gives spectral indices
that are close to zero, which is significantly different from what is
seen in the southern jet.  We do not consider this to be very likely.
Fitting for $f$ gives values in the range 0.98-0.995 --- allowing less
than 2\% of the light to pass unabsorbed would give a near perfect
fit.  Probably more likely is that there are variations in the
absorption over each beam area that account for the slightly flatter
than a pure free-free spectrum.  A detailed model has not been
attempted and almost surely would not be unique.

We are assuming that the origin of the jets, presumably in the
vicinity of a massive black hole, is located at the northern end of
the bright, compact regions seen at high frequency (Dhawan,
Kellermann, and Romney \markcite{D98}1998).  We believe that this is
consistent with the structures seen in our images, including the
absence of any compact features farther north in the high resolution
versions of the 43 GHz images.  But we cannot completely exclude the
possibility that the true base of the jets is significantly north of
this point as suggested by Nesterov, Lyuty, \& Valtaoja
\markcite{N95}(1995) based on time lags between optical and radio
variations.  If the core is displaced northwards, the lack of strong
radio emission at its location could be because the radio emission has
not turned on in the inner jet, as suggested by Nesterov, Lyuty, \&
Valtoaja, or because those inner regions are absorbed.  The latter
could be consistent with the fact that we do see structures in that
region at the high frequencies, but not at lower frequencies, and with
the possibility that the inverted spectrum in the region we call the
core is due to absorption.  A displaced core does not strongly affect
the conclusions of this paper, but would affect the parameterization
of the radial gradient of absorption.  It would also affect the
geometry derivation of WRB in the sense that the jets would lie closer
to the line-of-sight and the simple relativistic beaming model would
not work quite as well.

Our fit results show that the absorption has a strong gradient
away from the core.  This is seen clearly in the slices of
Figures~\ref{sliceJ} and \ref{sliceO} and in the gray scale images of
Figures~\ref{mapEMJ} and \ref{mapEMO}.  An attempt to characterize
this gradient was made using the slice results by fitting a power law
in $r_{2.5}$, the core distance in units of 2.5 pc.  This reference
distance was chosen so that the constant in the equation corresponds
to the value at about the middle of the northern feature.  The results
are displayed on Figures~\ref{sliceJ} and \ref{sliceO}.  They are:
\begin{equation}
L_{pc} ~n_e^2~ T_4^{-1.5}g_4 = 5.7 \times 10^8 r_{2.5}^{-1.5}
\label{tauJ}
\end{equation}
in January and 
\begin{equation}
L_{pc} ~n_e^2~ T_4^{-1.5}g_4 = 4.8 \times 10^8 r_{2.5}^{-1.8}
\label{tauO}
\end{equation}
in October.  The power law is clearly only a rough description of the
results.  The formal fit errors for the exponent are less than 0.1.
But it is clear that there are large deviations from any power law, so
the value of the exponent should not be taken too seriously.  A wider
range of core distances would be useful for any characterization of
the gradient.

The feature (NN), seen at about 80 mas north of the core at low
frequencies by STV, also has a peaked spectrum that STV conclude is
the result of free-free absorption.  For this feature, the absorption
is much weaker since it occurs at a much lower frequency.  The simple
power laws fit to our higher frequency data in the region less than 10
mas from the core, if extrapolated to the more distant feature, would
give far too much absorption.  If the power law characterization is
reasonable, the exponent must be in the vicinity of the $-2.6$
determined by STV.  But a power law with a $-2.6$ exponent is not very
consistent with our data in the 4 to 10 mas range.  It is likely
that the amount of ionized gas eventually falls off faster between 10
and 80 mas (3.2 and 26 pc) than it does over the 4 to 10 mas (1.3 to
3.2 pc) range over which we measure it.  In fact, the existence of
structure even closer to the core at 15 and 22 GHz (see
Figure~\ref{fullO}) suggests that the absorption does not continue to
increase as fast with decreasing core distance inside the region of
our measurements as it does further out.

\section{Discussion}

   The current data provide more frequencies than were available for
previous discussions of the spectrum of the northern feature and
provide measurements of all frequencies taken at approximately the
same time, so that time variability should not be a source of
uncertainty.  Based on the earlier 8.4 and 22 GHz data, with a 2 year
separation, VRB and LLV argue that the inverted spectrum of the
northern feature is not synchrotron self-absorption on the grounds
that, given the spectrum and the size of the feature, the magnetic
field energy density would have to be greater than the particle energy
density by a factor of $10^{14}$, which is unlikely.  Also the total
energy content of the jet with the implied magnetic field would be
unreasonably high.  The current measurements provide a rather simpler
argument.  The observed spectral index on the northern feature between
5.0 and 8.4 GHz is about $\alpha = 4$.  This is significantly steeper
than the optically thick synchrotron spectral index of 2.5.
Therefore, synchrotron self-absorption cannot be the only cause of the
inverted spectrum.  LLV also consider stimulated Raman scattering as
the cause of the inverted spectrum, but rule it out on the grounds
that the brightness temperature is inadequate.  That argument remains
unchanged.  The Razin-Tsytovich effect, which exponentially suppresses
synchrotron radiation at frequencies below those where the phase
velocity is greater than c, can cause a steeply inverted spectrum below $\nu_r
\simeq 20 n_e / B $ Hz.  But with only $\nu^{-1}$ in the exponent,
compared to the $\nu^{-2}$ of free-free absorption, it cannot produce
the change from about $\alpha = 0$, seen between 15 and 22 GHz, to
about $\alpha = 4$, seen between 5.0 and 8.4 GHz, without an
unphysically steep underlying emitted spectral index.  Also, the
Razin-Tsytovich effect operates in the emitting region rather than
along the line-of-sight, so it is difficult to understand why there
would be a very big difference between the northern and southern
features.  Free-free absorption, on the other hand, fits the spectra
reasonably well, and, to the extent that there are deviations, they
are what might be expected from non-uniformities in the medium,
something that is reasonable to expect.

   The opacity fits presented above give an emission measure of about
$5 \times 10^8$ at 2.5 pc with a fairly steep radial gradient,
assuming a temperature of $T=10^4$K.  The path through the absorbing
medium cannot be significantly larger than the projected distance of
the line-of-sight from the core.  In fact, if the geometry were that
of a thin disk, the path would be significantly smaller than the
projected offset.  The density of ionized gas required is roughly $n_e
= 2 \times10^4$ cm$^{-3}$ for a thick (1 pc) medium at $10^4$ K, rising
to significantly higher values for a thinner medium or a higher
temperature.

LLV point out that the Thomson optical depth along the
line-of-sight must not significantly exceed unity, constraining the
density to be 
\begin{equation}
n_e \lesssim 5 \times 10^5 L_{pc}^{-1} {\rm cm}^{-3}
\label{thomson}
\end{equation}
This density is not all that far from that implied by the free-free
absorption measurements.  If we insert the inequality of 
Equation~\ref{thomson} into Equations~\ref{tauJ} and \ref{tauO}, 
we get: 
\begin{equation}
T \lesssim 5.8 \times 10^5 r_{2.5} L_{pc}^{-2/3}
\end{equation}
for January and
\begin{equation}
T \lesssim 6.5 \times 10^5 r_{2.5}^{1.1} L_{pc}^{-2/3}
\end{equation}
for October.  The fact that $L_{pc}$ is not well constrained leaves
room for a fairly wide range of temperature limits.  But for values of
$L_{pc}$ between 0.01 and 1, the temperature at 2.5 pc cannot exceed
between about $1.4\times 10^7$ and $6.5\times 10^5$ respectively in
October and slightly less in January.  These limits don't really cause
any problems for models that we are aware of, but it is clear that the
temperature cannot greatly exceed the $10^6$ K called for in some
cases (eg. K\"onigl and Kartje \markcite{K94}1994; Gallimore, Baum, \&
O'Dea \markcite{G97}1997)

The region where free-free absorption is observed is a few parsecs
from the central object in NGC~1275.  LLV estimate the bolometric
luminosity to be about $4 \times 10^{44}$ ergs s$^{-1}$, which is
about the Eddington luminosity for a $10^6 M_{\sun}$ black hole.  If
that is the mass of the central object, a location 2.5 pc away is
at roughly $3 \times 10^7 R_G$ (The gravitational radius $R_G = 2 G M /
c^2$).  On the other hand, the same location would only be about $3
\times 10^4 R_G$ from a $10^9 M_{\sun}$ black hole.  These distances
are in the outer regions typically covered in accretion disk models,
if not well beyond.  In such regions, models are complicated by the
possible importance of self gravity and of external illumination of
unknown geometry.  The temperature at these distances is expected to
be too low for ionization without illumination by ionizing radiation
from the central regions.  Such illumination requires an appropriate
geometry such as a flaring or warped disk or a source of radiation
offset from the disk center, possibly from a jet or from scattering.
LLV give an extensive discussion of the implications for disk models
of the observation of free-free absorption in NGC~1275 --- a
discussion that is not significantly changed by the new data.

The idea that AGN disks are likely to be illuminated by ionizing
radiation from the central compact region near the black hole has been
explored as part of efforts to model the spectra of broad line regions
and the structure of outer disks (see Collin and Hur\'e
\markcite{C99}1999 and references therein including Collin-Souffrin \&
Dumont \markcite{C90}1990).  Generally the density of the outer disk
is expected to be much higher than implied by the free-free
absorption.  However, much of the disk is sufficiently optically thick
that the central regions are shielded from ionizing radiation.  These
regions will be neutral and will not contribute to the free-free
absorption or to many of the observed optical emission lines.  But
there is likely to be an ionized ``chromosphere'' above the dense
regions.  At large radii, the density is too low to provide shielding
and the disk is fully ionized and at a temperature of about 7000 K.
This chromosphere, or the ionized regions at large distances, could be
the absorbing region.  

An alternative to the irradiated disk for the location of the ionized
material responsible for the free-free absorption might be a wind off
the surface of the disk.  For example, K\"onigl and Kartje
\markcite{K94}(1994 --- see also other references therein) discuss the
properties of centrifugally driven winds in the AGN context.  Winds of
this type are the leading model for the origin of the bipolar outflows
that are very common in young stellar objects.  They seem likely to
exist whenever a disk is threaded by magnetic fields --- a situation
that is probably hard to avoid.  The temperatures and densities of
some of the examples presented in K\"onigl and Kartje are of the
magnitude required by our data.

It seems likely that both of the above scenarios apply in an AGN.  The
central regions are a strong source of ionizing radiation which is
very likely to affect some portion of the disk.  And, given the almost
certain presence of magnetic fields in any accretion disk, combined
with the ubiquitous presence of magnetically driven winds from
galactic examples of accretion disks, it seems hard to avoid having a
magnetically driven wind.  There is a considerable range of possible
variations on these models, depending on the black hole mass, the
accretion rate, the disk composition, the magnetic field strength, and
other factors, so a variety of types of data will be required to
determine anything like a unique model.  Because of the large range of
possible models, we have not attempted detailed comparisons with any
specific models.  But our data do provide some firm constraints that
any model must match.  We are attempting to obtain additional
constraints by making multi-frequency observations that can be used at
higher resolution than those presented here.  Such observations will
also be sensitive to temporal variations in the absorption since 1995.
In addition, we are searching for recombination lines from the
absorbing medium.

\section{Conclusions}

The primary conclusions of this work are:

\begin{itemize}

\item The suggestion from VRB and WRB, based on two frequencies and
observations separated by 2 years, that there is free-free absorption
of the northern feature in 3C~84 at a few parsecs from the central
object, is confirmed.  We present two separate epochs in which the
absorption was observed nearly simultaneously at 5 frequencies.  The
observed spectral indices are sufficiently steep to preclude other
absorption mechanisms.

\item The free-free absorption shows a two dimensional structure
dominated by a gradient with distance from the radio core.  The
absorption is greater near the core and falls off with distance.  If
the absorption goes as a power law with distance, the exponent is a
bit above $-2$ over the range of the observations presented here
(about 1.5 to 3.5 pc projected distance from the core).  However a
somewhat steeper exponent is required to match up with the
observations of STV at around 25 pc.  Those observations were taken at
the same time as ours.

\item The observed absorption is consistent with the model proposed by
WRB and VRB.  In that model, the northern feature is on the far
side of the system relative to the Earth.  There is an accretion disk
extending to the parsec scales observed here and that accretion disk
has associated ionized gas that is responsible for the absorption.
The amount of ionized gas falls off fairly rapidly with core distance.

\end{itemize}

Various models, including disk ionization by radiation from the
central regions and disk-driven hydromagnetic winds, might provide the
necessary ionized material in a geometry that would only affect the
far-side jet.  The free-free absorption results provide firm
constraints on the ionized gas on parsec scales, including positional
information, that any model of the central regions of NGC~1275 must
match.

\acknowledgements 

We would like to thank J. Benson and W. Alef for their contributions
to these observations.  We also thank J. Wrobel and G. Taylor for
useful discussions.  Finally, we thank the staff of the VLBA for their
invaluable support.  These observations would not have been possible
without the major advances in frequency flexibility and image quality
provided by the VLBA.  The National Radio Astronomy Observatory is a
facility of the National Science Foundation, operated under
cooperative agreement by Associated Universities, Inc.

\end{document}